\documentclass[aps, pra, twocolumn, superscriptaddress, groupedaddress]{revtex4-1}
\usepackage{calc}
\usepackage{array}
\usepackage{natbib}
\usepackage{siunitx}
\usepackage{hyperref}
\usepackage{graphicx}
\usepackage{mathtools}
\usepackage{amsmath, amssymb, amsbsy, bm}
\usepackage{xcolor}

\DeclarePairedDelimiter\abs{\lvert}{\rvert}%

\begin{document}

\title{Effects of disorder upon transport and Anderson Localization\\
       in a finite, two dimensional Bose gas}

\author{Mojdeh S. Najafabadi}       \email{shimo048@postgrad.otago.ac.nz}
\author{Daniel Schumayer}           
\author{David A.~W. Hutchinson}     
\affiliation{Dodd-Walls Centre for Photonic and Quantum Technologies,
             Department of Physics, University of Otago,
            Dunedin 9016, New Zealand}

\begin{abstract}
    Anderson localization in a two-dimensional ultracold Bose-gas has been demonstrated experimentally. Atoms were released within a dumbbell-shaped optical trap, where the channel of variable aspect ratio provided the only path for particles to travel between source and drain reservoirs. This channel can be populated with columnar (repulsive) optical potential spikes of square cross section with arbitrary pattern. These spikes constitute impurities, the scattering centres for the otherwise free propagation of the particles. This geometry does not allow for classical potential trapping which can be hard to exclude in other experimental setups. Here we add further theoretical evidence for Anderson localization in this system by comparing the transport processes within a regular and a random pattern of impurities. It is demonstrated that the transport within randomly distributed impurities is suppressed and the corresponding localization length becomes shorter than the channel length. However, if an equal density of impurities are distributed in a regular manner, the transport is only modestly disturbed. This observation corroborates the conclusions of the experimental observation: the localization is indeed attributed to the disorder. Beyond analysing the density distribution and the localization length, we also calculate a quantum `impedance' exhibiting qualitatively different behaviour for regular and random impurity patterns.
\end{abstract}

\date{\today}
\maketitle

\section{Introduction \label{sec:introduction}}

All real medium, however pure, contain disorder which influences its transport properties \cite{Haug1972, Solyom2007, Authier2013, Lee1985, ziman1797disorder}. Indeed, disorder is essential for transport in a regular lattice, for otherwise particles undergoing Bloch oscillations \cite{Ashcroft1976} and are localised. However, impurities (interstitial atoms, lattice defects, etc), or even the finite size of the system, which destroy the perfect spatial periodicity, also give rise to residual resistance against electron flow \cite{Ziman1960}. If the density of impurities is high enough and the typical electron energy is low, the electrons are localized \cite{Anderson1958, Mott1960, Mott1961, Stollmann2012}, hence the substance is an insulator. The opposite, the absence of translational symmetry alone, however, does not guarantee an insulating phase \cite{Shechtman1984, Sokoloff1987, Sokoloff1988, Jian1992}. The role of disorder in transport can therefore be very subtle.. 

One striking effect of random disorder is the suppression of transport due to destructive interference via multiple propagation paths, and the consequent confinement of wave packets. This phenomenon is known as Anderson localization  \cite{Anderson1958, Abrahams1979, Abrahams2010}. This single particle wave phenomenon does not require any special interaction between particles or specific geometry and so appears ubiquitously in nature and can be present in all kinds of systems \cite{Mott1969,Lee1985, Weaver1990, McCall1991, Wiersma1997, Genack1997, Weiland1999, Storzer2006, Laurent2007, Schwartz2007, Hu2008, Chabe2008, Riboli2011, Sperling2012, Lopez2012, Manai2015, Ying2016}. Since Anderson's early proposal, the effects of interaction \cite{Fishman2012, Shepelyansky1993}, dimensionality \cite{Abrahams1979}, violation of time-reversal symmetry \cite{Bergmann1984}, and spin-orbit coupling \cite{Bergmann1982} upon localization have all been analysed. 
\begin{figure}[b]
    \includegraphics[width=55mm]{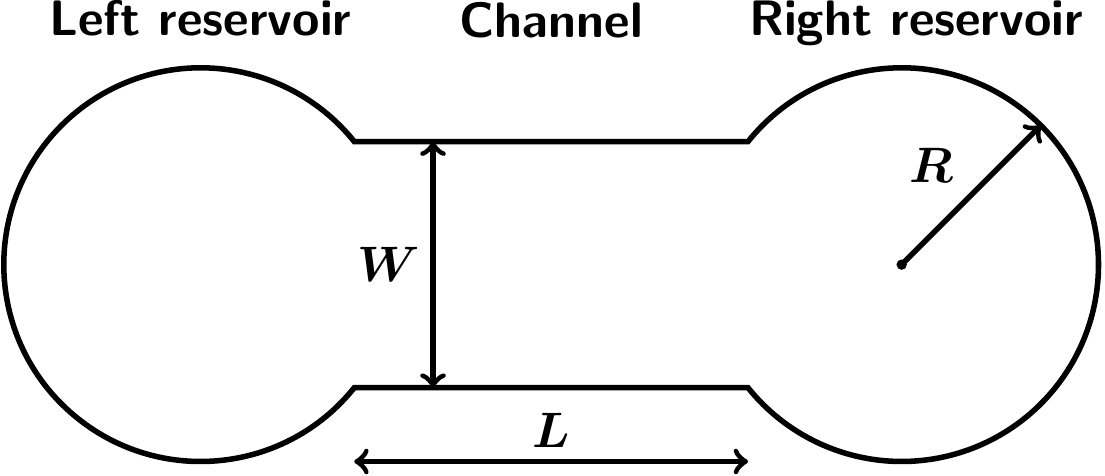}
    \includegraphics[width=27mm]{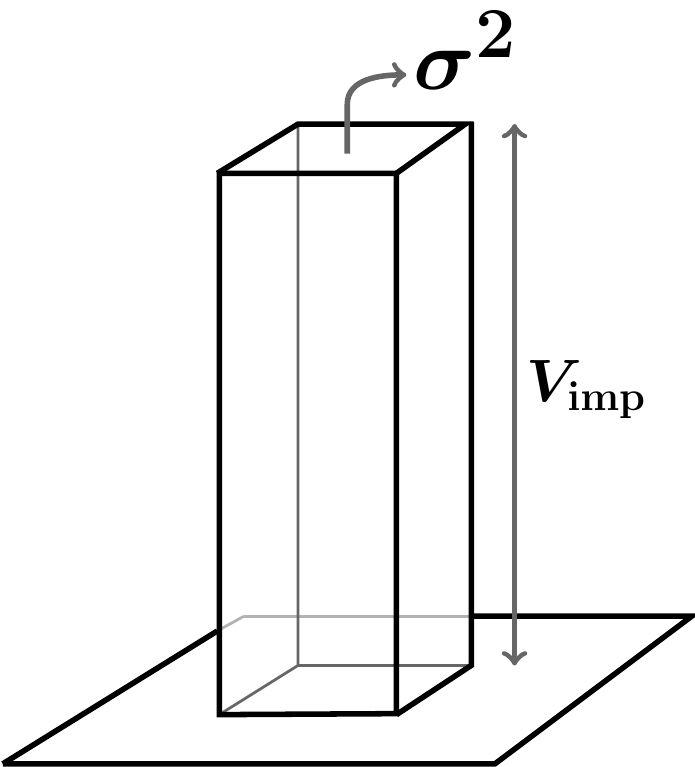}
    \caption{\label{fig:DumbbellSchematics}
             On the left the schematics of the dumbbell potential is shown: $L$ is the length of the channel, $W$ is the width of the channel, while $R$ is the common radius of the two circular reservoirs. On the right the single columnar potential spike is depicted, showing its square cross section of area $\sigma$ and height/strength of $V_{\text{imp}}$.
            }
\end{figure}

Ultracold atomic systems, with their experimental flexibility and precision control of both internal and external degrees of freedom, make them an excellent platform to explore Anderson localization \cite{Kondov2011, Billy2008, Roati2008, deMarco2011, Aspect2012}. Localization of atoms have been demonstrated in 1D \cite{Billy2008, Roati2008}, and also in 3D \cite{deMarco2011, Aspect2012} where, for the latter, the localized and delocalized states are separated by a well-defined mobility edge \cite{Mott1975}. However, observation of Anderson localization in two-dimensional ultracold systems proved to be elusive \cite{White2020}. Although weak localization in 2D has been reported \cite{Vincent2010, Jendrzejewski2012}, the relatively high percolation threshold of a speckle potential posed serious difficulty and resulted in classical localization. In this regard \citet{Morong2015} proposed a realistic point-like disorder potential circumventing the problem of percolation limit. In 2020 \citet{White2020}, after designing a highly flexible optical setup overcoming these technical challenges, demonstrated Anderson localization in a 2D ultracold system. 

Here we provide further computational analyses of localization extending this previous work \cite{White2020}. We consider a 2D dumbbell-shaped trap with potential-spikes distributed within the channel. An interacting Bose-Einstein condensate is prepared in its ground state within an initial harmonic trap and centered at the middle of the channel. The harmonic trap is then turned off and the condensate is released within the dumbbell-shaped trap. Although the self-interaction is taken into account in the simulation, mainly to remain close to real-life experiments, its effect is only observable at the very beginning of evolution, when the mean-field energy drives the expansion within the channel. As the density falls rapidly the interaction together with its influence on the localization diminishes swiftly.

Our main focus is on the transport of particles in two scenarios: impurities distributed randomly or regularly within the channel. We expect these two scenarios to exhibit fundamentally different transport properties. In the analyses we rely on three quantitative measures: the localization length, $\xi$, the momentum distribution, $\mathbf{k}$, and on an atomtronic impedance, $Z$, introduced in analogy with that of electrical impedance.

\section{Details of simulations}

After introducing scales for the energy, and for the temporal and spatial coordinates as $\hbar \omega_{0}$, $\omega_{0}^{-1}$, $\sqrt{\hbar/m \omega_0}$, respectively, one obtains the numerically more convenient form of the Gross-Pitaevskii equation
\begin{equation} \label{eq:DimensionlessGPE}
    i \frac{\partial \psi}{\partial t}
    =
    \left \lbrack 
        -\frac{1}{2} \nabla^{2} +  U + \beta \abs{\psi}^{2}
    \right \rbrack
    \psi.
\end{equation}
The time evolution of the condensate at absolute zero temperature is described by this nonlinear equation adequately. Here $U$ is the dimensionless potential including two terms: the dumbbell trap, $V_{\text{db}}$, and the overall sum of potential spikes, $V_{\text{disorder}}$, representing the impurities. As in Ref.~\cite{White2020}, the strength of impurities, $V_{\text{imp}}$, is constant together with the footprint $\sigma^{2}$ of a single spike (see Fig.~\ref{fig:DumbbellSchematics}). The mean-field interaction is measured by $\beta = 2 \sqrt{2} \pi N a_{s}/a_{z}$, with $a_{z}$ being the oscillator length corresponding to $\omega_{z}$, $N$ is the number of \textsuperscript{87}Rb atoms, and $a_{s}=107 a_{0}$ is the $s$-wave scattering length. In Eq.~\eqref{eq:DimensionlessGPE} $\psi$ is normalised to unity. Parameter values are in Table~\ref{tab:NumericalParametes}.

\begin{table}[t]
  \begin{tabular}{p{43mm}p{1mm}lp{1mm}l}
  Description                       & &                  & & Value                           \\
  \hline\hline
  Spatial extension ($x$ direction) & & $L_{x}$          & & 500\,\si{\micro\meter}          \\
  Spatial extension ($y$ direction) & & $L_{y}$          & & 225\,\si{\micro\meter}          \\
  Grid points ($x$ direction)       & & $n_{x}$          & & 1536                            \\
  Grid points ($y$ direction)       & & $n_{y}$          & & 768                             \\
  Reservoir radius                  & & $R$              & & 43.2\,\si{\micro\meter}         \\
  Channel length                    & & $L$              & & 180\,\si{\micro\meter}          \\
  Channel width                     & & $W$              & & 36\,\si{\micro\meter}           \\
  Dumbbell potential depth          & & $V_{\text{db}}$  & & \SI{52}{\nano\kelvin}           \\
  Impurity height                   & & $V_{\text{imp}}$ & & \SI{17}{\nano\kelvin}           \\
  Impurity cross-section area       & & $\sigma^{2}$     & & \SI{1.4}{\micro \meter\squared} \\
  Number of particles               & & $N$              & & 16000                           \\
  Frequency of harmonic trap        & & $\omega_{0}$     & & $2\pi \times$\,\SI{25}{\radian\per\second} \\
  Frequency of squeezing trap       & & $\omega_{z}$     & & $2\pi \times$\,\SI{800}{\radian\per\second} \\
  Bohr radius                       & & $a_{0}$          & & $5.29 \times 10^{-11}$\,\si{\meter} \\
  Range of fill-factor              & & $\eta$           & & $[0, 0.3]$
  \end{tabular}
  \caption{\label{tab:NumericalParametes} Symbols and values of parameters.}
\end{table}

The dumbbell-shaped trap, depicted in Fig.~\ref{fig:DumbbellSchematics}, consists of two reservoirs separated by a channel with point scatterers distributed randomly or regularly within the channel. The simulation starts with calculating the ground-state of the condensate trapped by the harmonic trap using the imaginary time propagation method \cite{Kosloff1986}. This stationary solution serves as the initial condition for the time-evolution, and the condensate is allowed to expand and propagate within the channel, decorated with impurities, towards the reservoirs. Equation \eqref{eq:DimensionlessGPE} is solved numerically by the Runge-Kutta-Fehlberg method \cite{Press1992}. 

\begin{figure}[t!]
    \includegraphics[width=70mm]{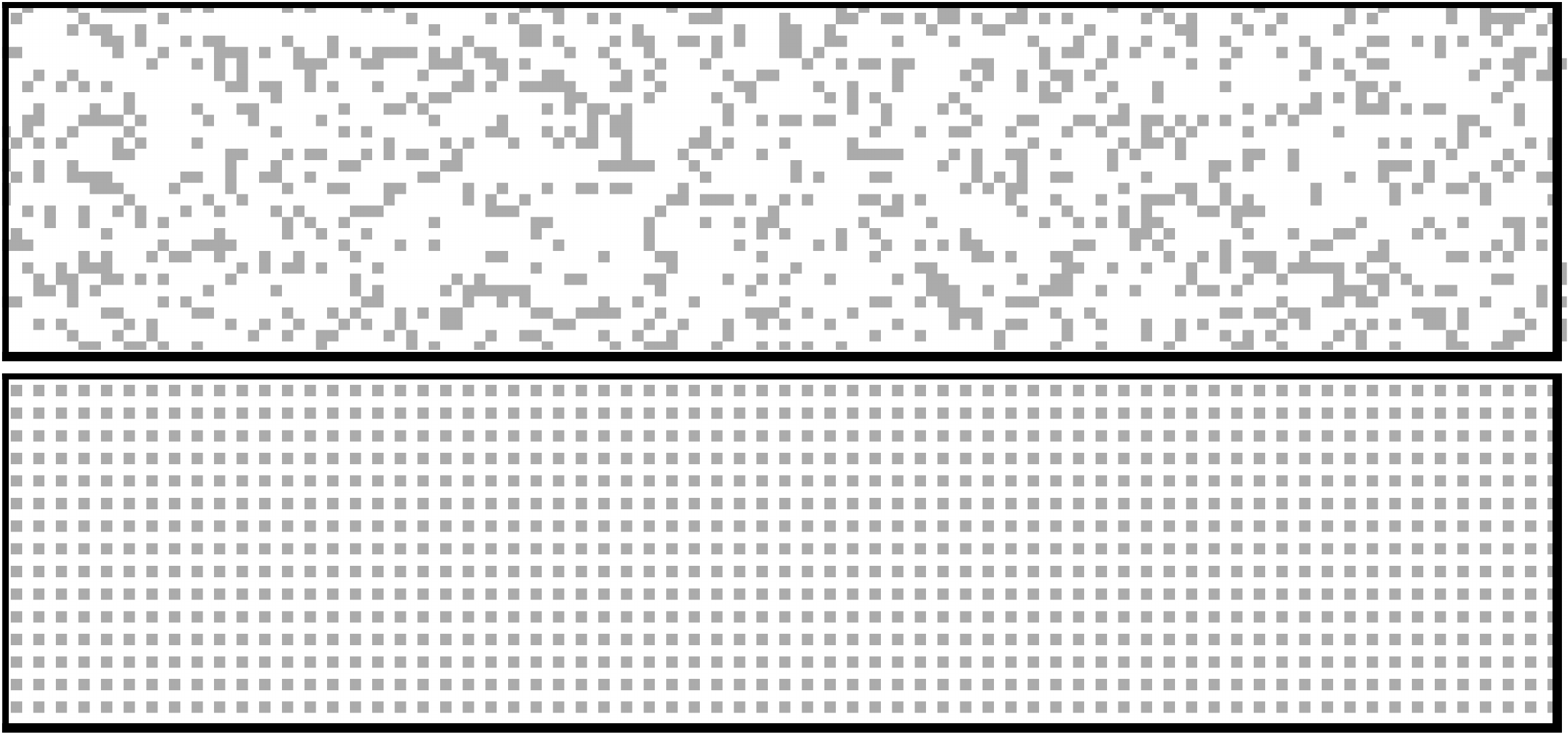}
    \caption{\label{fig:RegularlyAndRandomlyDistributedImpurities}
             Impurity potential, $V_{\text{imp}}$, is depicted for randomly (top) and regularly (bottom) distributed potential spikes for the same fill-factor $\eta=0.25$. Only the channel segments of the dumbbell trap are shown.
            }
\end{figure}

In order to quantify the ``amount'' of disorder in the channel we introduce a geometric measure, $\eta$, based on the overall footprint of the impurities relative to the total available area within the channel
\begin{equation}
    \eta = \frac{A_{\text{disorder}}}{A_{\text{channel}}} = n\sigma^{2},
\end{equation}
where $n$ is the density of scatterers and $\sigma^{2}$ is the cross-section area of a single scatterer. For geometrical reasons one may call $\eta$ as fill-factor. Its meaning is apparent in Fig.~\ref{fig:RegularlyAndRandomlyDistributedImpurities} where the channel segments of the dumbbell are shown for $\eta=0.25$ with the impurities distributed randomly (top) or regularly (bottom).

We estimate the localization length, $\xi$, from fitting exponential decay on the two- and one-dimensional probability densities, defined as $\rho_{\text{2D}} = \abs{\psi}^{2}$ and
\begin{equation*}
    \rho_{\text{1D}}(x)
    =
    \int{\!\rho_{\text{2D}} (x, y) dy}
    =
    \int{\abs{\psi(x,y)}^{2} dy},
\end{equation*}
or in other words, the one-dimensional density is the one-dimensional column density along the $x$ axis, usually easily accessible in an experimental setup.

\section{Results}

\begin{figure}
    \includegraphics[width=82mm]{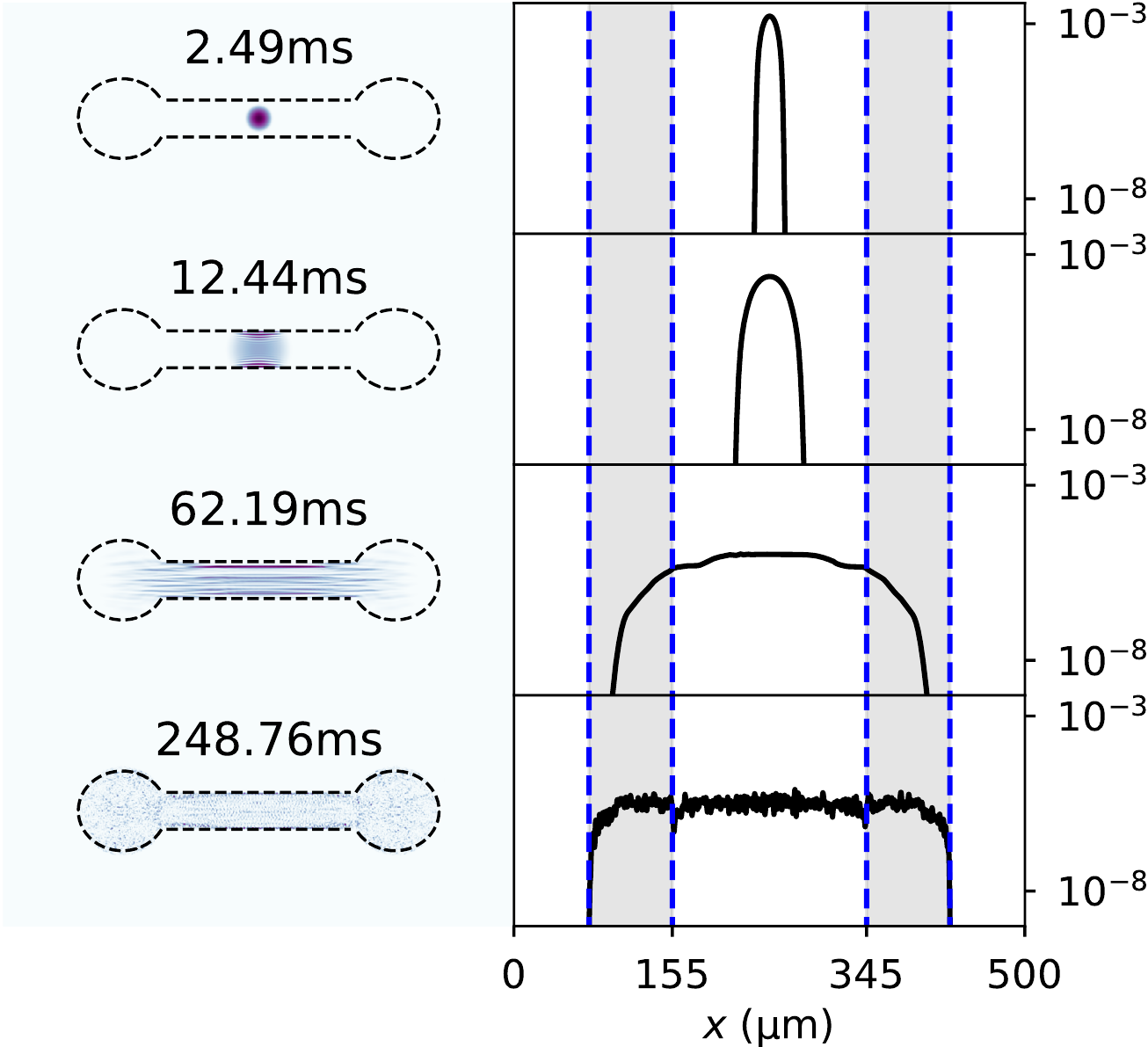}
    \caption{\label{fig:SnapshotsOfTimeEvolutionForZeroEta}
             The left and right columns show $\rho_{\text{2D}}$ and $\rho_{\text{1D}}$, respectively, at different moments in time covering the entire duration of simulation. In the right column $\rho_{\text{1D}}$, is plotted on a logarithmic scale, while the left and right reservoirs are depicted as shaded areas. The channel length is 180\,\si{\micro \meter}.
             While there is small portion of the density falling outside of the dumbbell potential, it does not show up in these graphs, as this portion is negligible, thus several order lower in magnitude than within the dumbbell.
            }
\end{figure}

Before we analyse the two main scenarios, let us demonstrate the time evolution of the condensate without potential-spikes. This case serves as a benchmark. Figure~\ref{fig:SnapshotsOfTimeEvolutionForZeroEta} shows both the one- and two-dimensional densities as the condensate expands within the channel in the absence of any impurity. It is clear that by $t \simeq$\SI{250}{\milli\second} the atoms fill the entire dumbbell trap more or less uniformly, exactly what one would expect for a gas trapped in a finite volume.
In the following two subsections we describe how the spatial distribution of potential spikes alter particle transport.

\subsection{Randomly distributed scatterers}

We study the long-time behaviour of $\rho_{\text{1D}}$ in the channel and in the two reservoirs. For high enough fill-factors we expect localization to occur, hence the density develops an exponentially decaying profile while expanding within the disordered dumbbell
\begin{equation}\label{fitting}
    \rho_{\text{1D}}(x)
    \propto
    \rho_{0}\, \exp{\!\left ( -\frac{2 \abs{x}}{\xi} \right )},
\end{equation}
where the origin is at the centre of the channel. We call the characteristic parameter $\xi$ localisation length.

\begin{figure}
     \includegraphics[width=82mm]{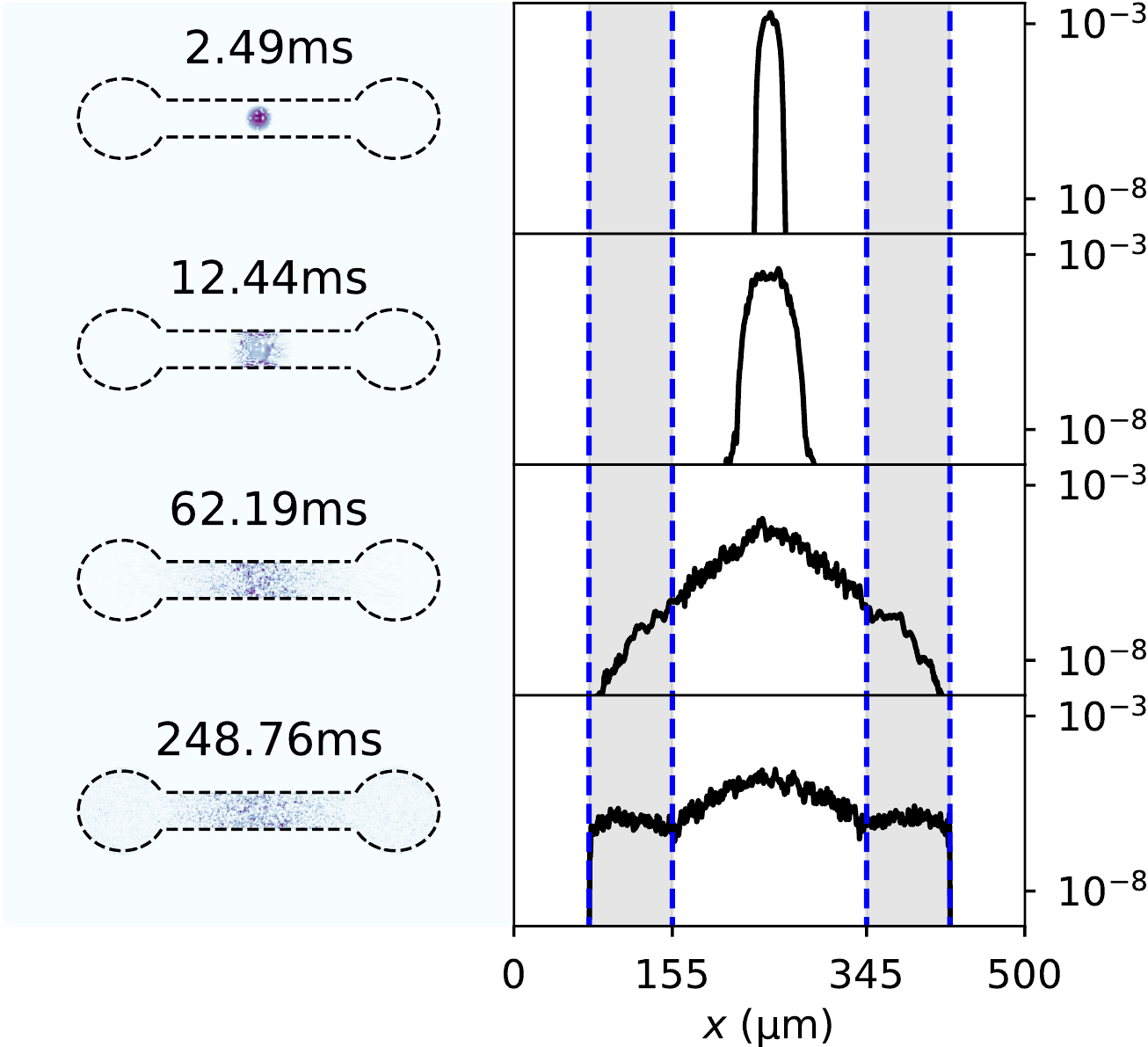}
     \includegraphics[width=82mm]{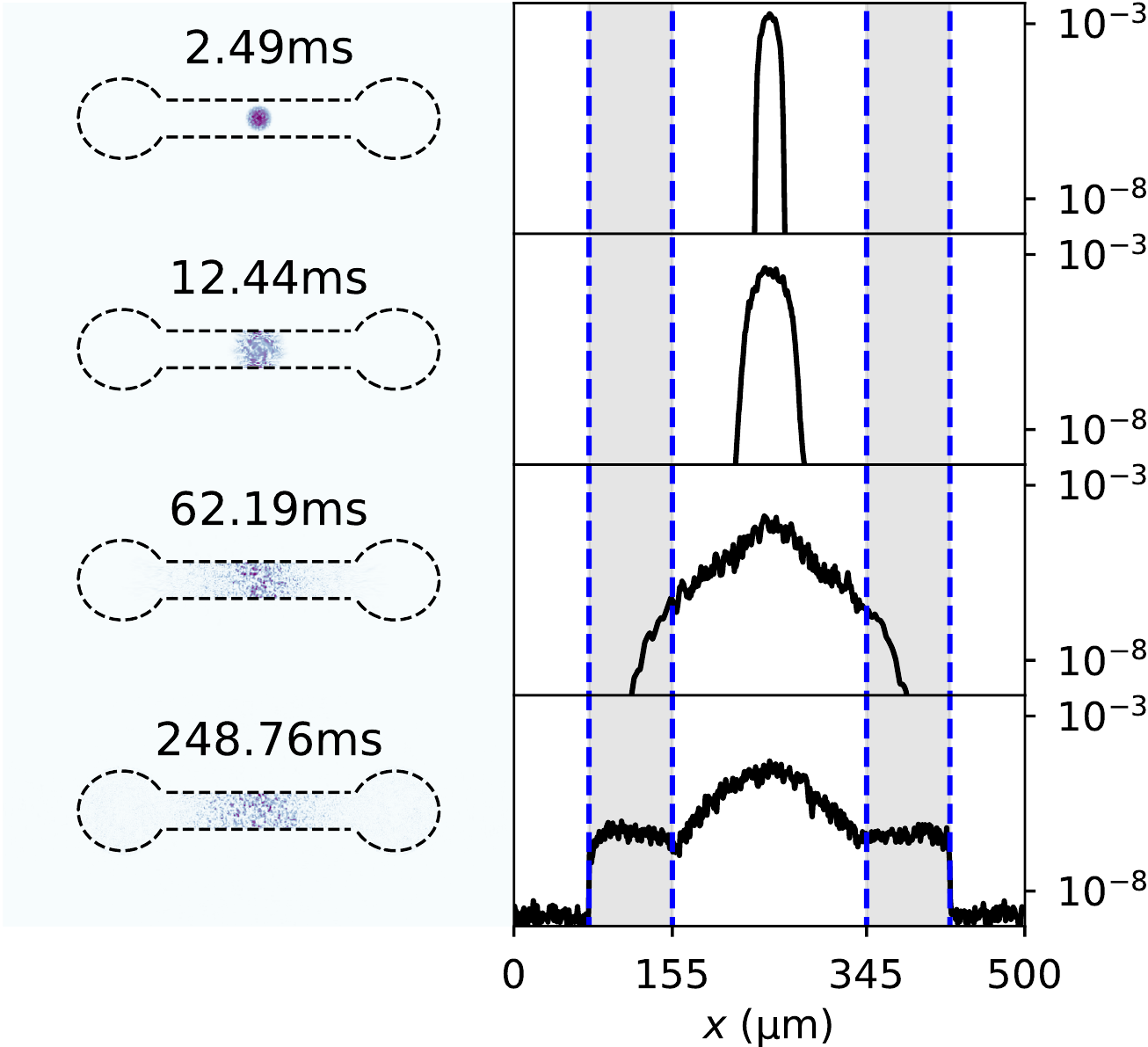}
     \caption{\label{fig:SnapshotsOfTimeEvolutionForHighEta}
              In the right column the time evolution of $\rho_{\text{1D}}$ is depicted for $\eta=0.05$ (top) and $\eta=0.2$ (bottom) at $t=2.49$, 12.44, 62.19 and at $t=248.76$\,\si{\milli\second}. In the left column $\rho_{\text{2D}}$ is shown with the contour of the dumbbell potential overlaid. The geometry is determined by $(L, W, R) = (180, 36, 43)$\,\si{\micro\meter}.
             }
\end{figure}

Figure~\ref{fig:SnapshotsOfTimeEvolutionForHighEta} shows $\rho_{\text{1D}}$ and $\rho_{\text{2D}}$ profiles at four moments in time as the condensate expands within the channel with different amount of disorders, $\eta = 0.05$ and 0.2. For small fill-factor $\rho_{\text{1D}}$ does not show any appreciable triangular shape at the centre excluding the possibility of localization. However, for higher disorder, a central peak develops.
Towards the end of time evolution the reservoirs also hold non-negligible amount of matter. These atoms escape from the disordered channel as their kinetic and mean-field energy is high enough. We will substantiate this claim later by calculating the momentum (and hence energy) distributions of particles in the reservoirs and in the channel. However, increasing $\eta$ further reduces the density in the reservoirs despite the fact that the strength of individual potential-spiked has not changed. 

In order to quantify localization Eq.~\eqref{fitting} is fitted to $\rho_{\text{1D}}$ for $t>200$\,\si{\milli \second} treating $\xi$ as fitting parameter. As $\log{\!(\rho_{\text{1D}})} \propto \abs{x}$ we defined $\xi_{\text{left}}$ and $\xi_{\text{right}}$ based on the density for $x<0$ and $x>0$.
\begin{figure}
     \includegraphics[width=82mm]{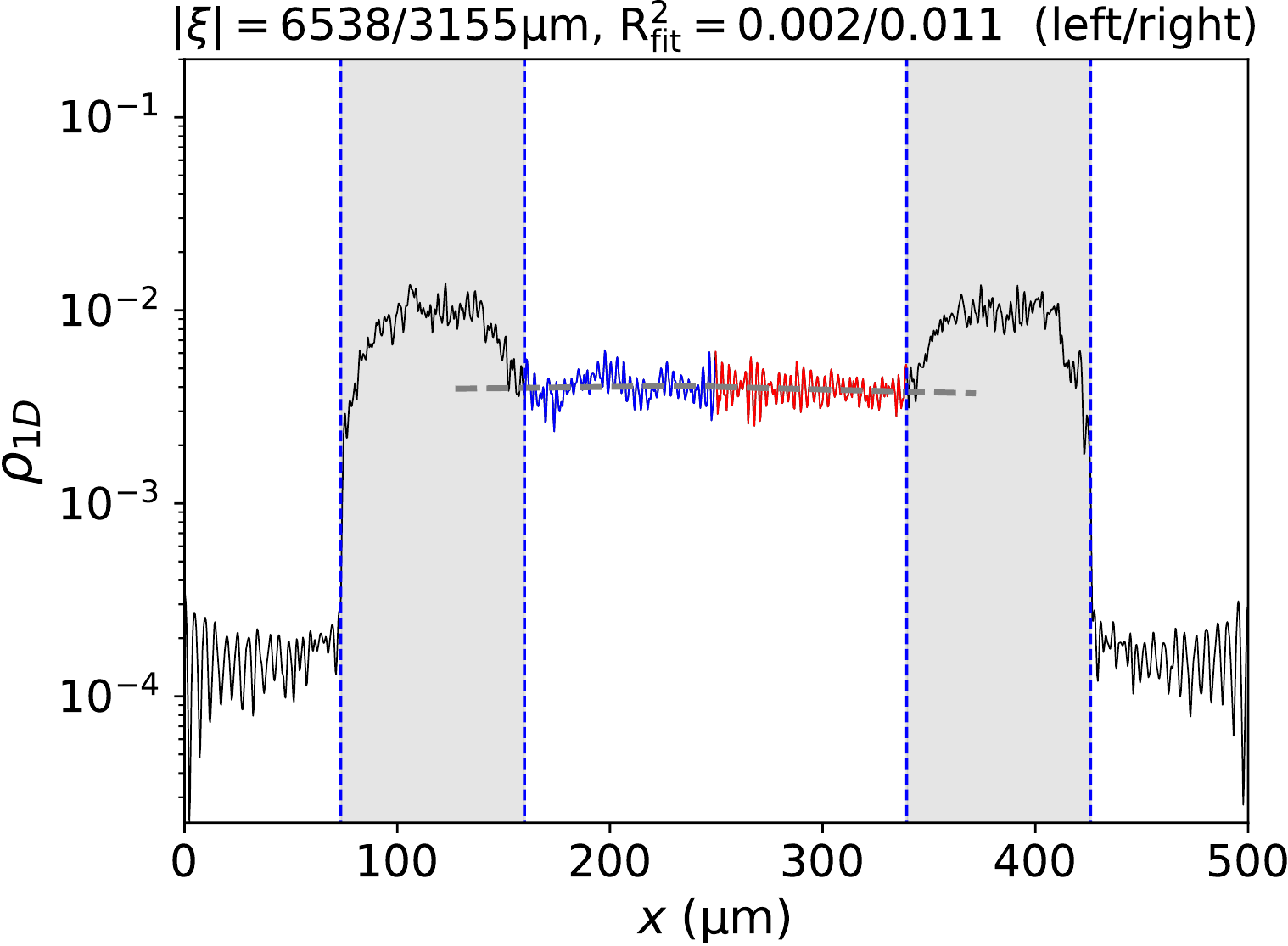}
     \includegraphics[width=82mm]{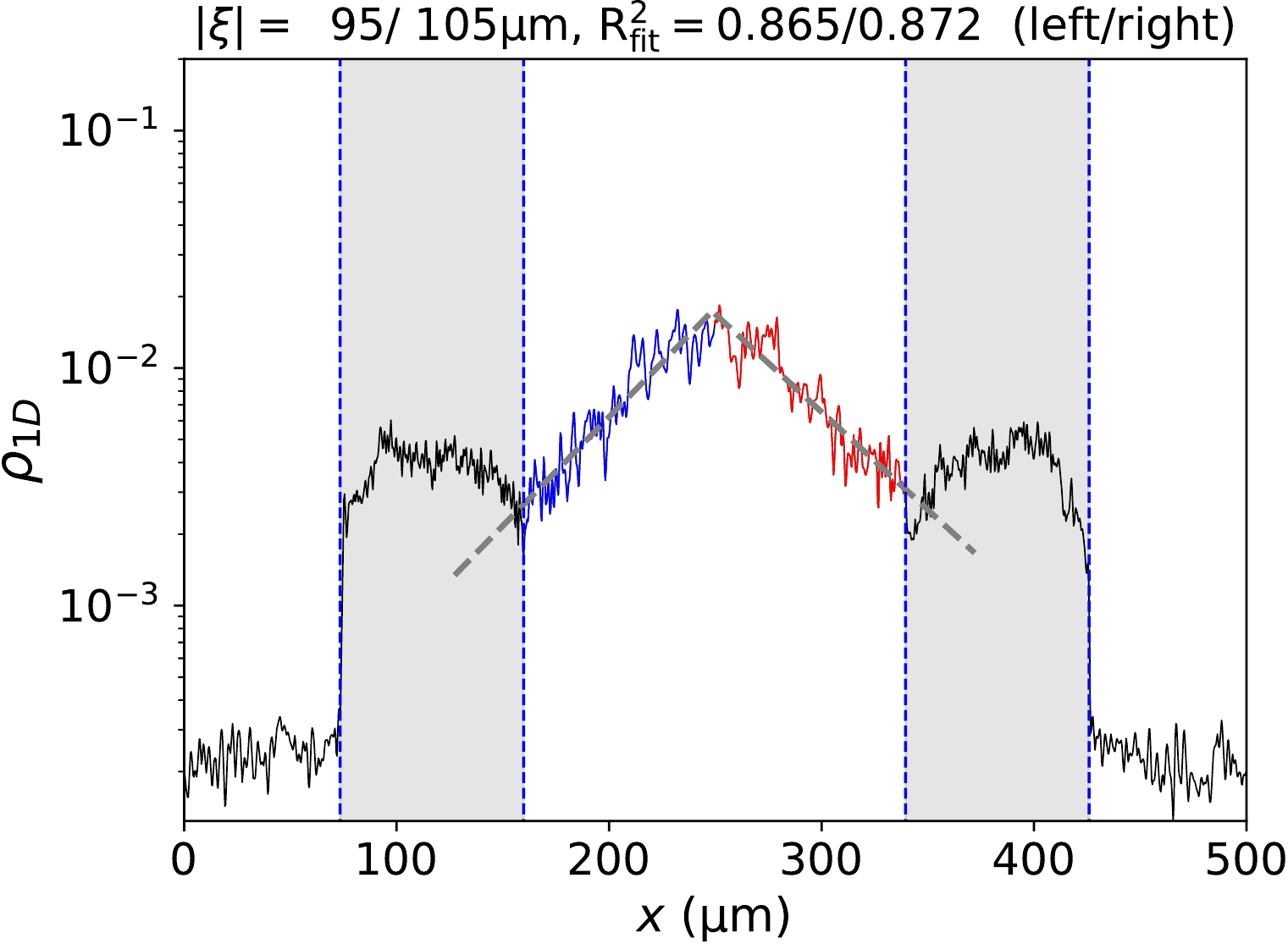}
     \includegraphics[width=82mm]{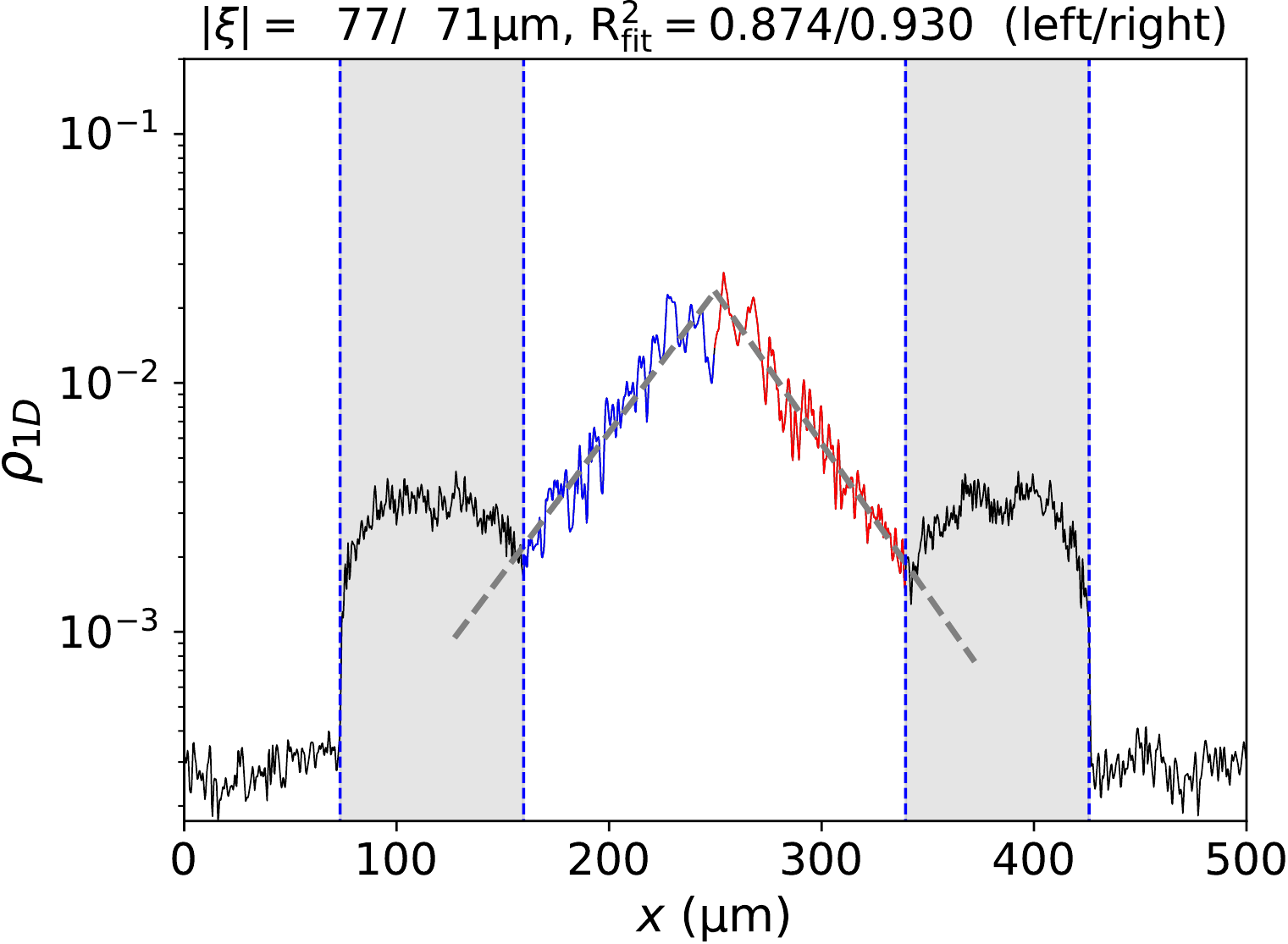}
     \caption{\label{fig:FittingExponentialDecayOn1DDensity}
              Figure depicts the one-dimensional density profile over the entire numerical box for $\eta=0$ (top), $0.1$ (middle), and $0.2$ (bottom). The vertical dashed lines separate the five important regions: the outermost regions are outside of the dumbbell trap, while the central three regions are the left reservoir, channel, and right reservoir. Within the channel the density profile is plotted using two colours for the left and right side of the channel, respectively. The graph also shows the linear fits to $\rho_{\text{1D}}$. The fitted $\xi_{\text{left}}$, $\xi_{\text{right}}$, together with the goodness-of-fit, $R_{\text{fit}}^{2}$ are given at the top of each graphs.
             }
\end{figure}
In Figure~\ref{fig:FittingExponentialDecayOn1DDensity} some of the curve-fittings are shown for $\eta=0$ (top), 0.1 (middle), and 0.2 (bottom) together with $\log{\!(\rho_{\text{1D}})}$ for $t > 200$\,\si{\milli \second}. The localization lengths $\xi_{\text{left}}$, $\xi_{\text{right}}$, and the goodness-of-fit, $R_{\text{fit}}^2$, are also provided above each panel \footnote{The goodness-of-fit, $R_{\text{fit}}^2$, varies in [0,1] and indicates the quality of the fit.}. The top graph of Fig.~\ref{fig:FittingExponentialDecayOn1DDensity} shows no exponential decay in either side of the channel, hence the goodness-of-fit is almost zero. In contrast, $\rho_{\text{1D}}$ in the middle panel of Figs.~\ref{fig:FittingExponentialDecayOn1DDensity} starts developing an exponential decay and consequently $R_{\text{fit}}^{2}$ is elevated to $\sim$0.87. As $\eta$ increases we expect more particles being localized within the channel due to more interference events, and as a result, $\xi$ decreases: at $\eta=0.1$ one finds $\xi_{\text{left}}/ \xi_{\text{right}} = 95/105$\,\si{\micro \meter}, while $\xi_{\text{left}}/ \xi_{\text{right}} = 75/70$\,\si{\micro \meter} for $\eta=0.2$. Note, as the impurities are distributed randomly, there is no equality in their number in each side of the channel. Therefore, $\xi$ in two sides of the channel can be slightly different. However, we expect that over numerous configurations at fixed $\eta$, this difference would vanish.

\begin{figure}
    \includegraphics[width=82mm]{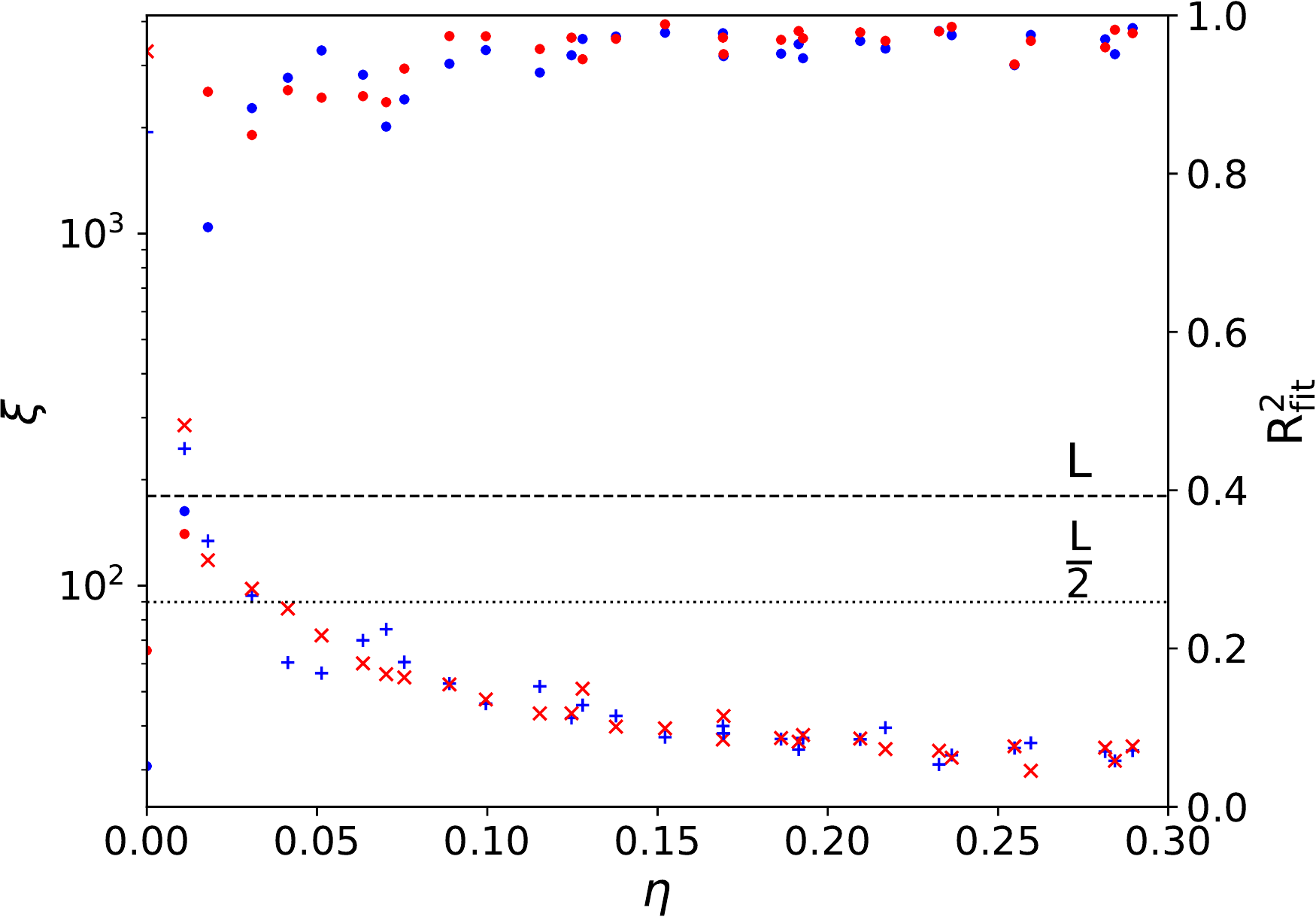}
    \caption{\label{fig:LocalizationLength_fillfactor}
             Localization lengths, $\xi_{\text{left}}$ and $\xi_{\text{right}}$, are depicted as functions of $\eta$ with blue and red crosses, respectively (left axis). The corresponding measure for goodness-of-fit, $R_{\text{fit}}^{2}$, is also shown with filled circles and with applying the same colour coding (right axis).
            }
\end{figure}

Figure~\ref{fig:LocalizationLength_fillfactor} depicts $\xi$ as a function of $\eta$. The localization lengths are calculated by taking an average over the last 50 snapshots of $\rho_{\text{1D}}$. The circular markers represent $R_{\text{fit}}^2$ and show an upward trend as the linear fit becomes better and better for increasing $\eta$. More importantly, we also see that $\xi$ falls below the half of the channel length for $\eta > 0.07$.

\begin{figure}
    \includegraphics[width=75mm]{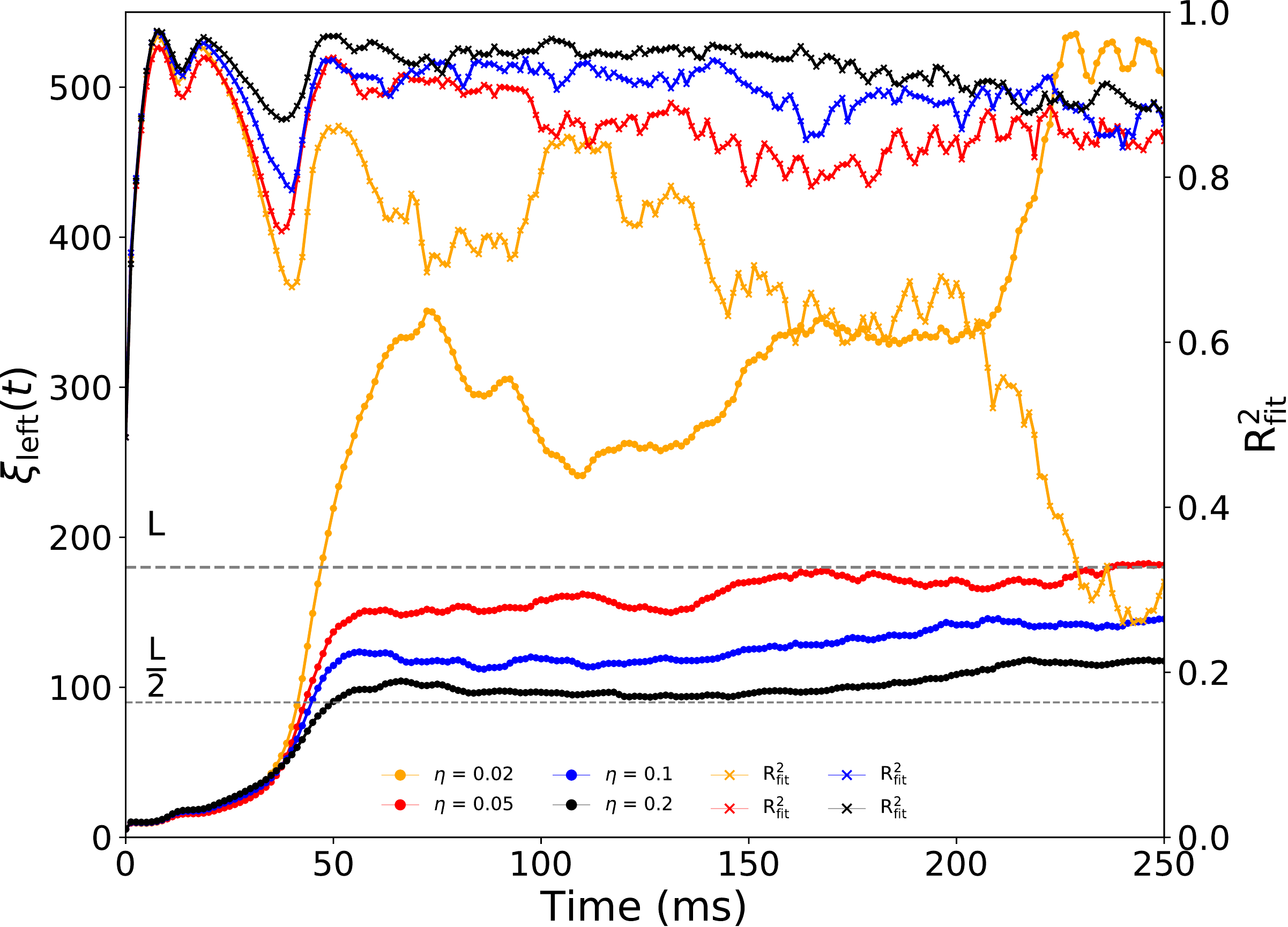}
    \caption{\label{fig:localization_vs_L0_time_L250}
             The apparent localization length, $\xi_{\text{left}}$, is drawn at each time step for $\eta=0.02$, 0.05, 0.1 and 0.2. The corresponding goodness-of-fit measures are also plotted with matching colour. The dumbbell geometry is given by $(L, W, R) = (180, 36, 45)$\,\si{\micro \meter}. The horizontal gray dashed lines represent the length of the channel, $L$, and its half, $L/2$, in order to provide comparison.
            }
\end{figure}

We may briefly scrutinise the dynamics of localisation as well. The localized particles have a fixed localization length unlike for particles in extended states. Figure~\ref{fig:localization_vs_L0_time_L250}, shows $\xi(t)$ and $R_{\text{fit}}^2(t)$ for four values of $\eta$ in $[0.02, 0.2]$. For low disorder atoms occupy extended states hence $\xi(t)$ increases and subsequently $R_{\text{fit}}^2$ decreases. For higher $\eta$ we see more stability in localization length as a function of time. One can argue for a slightly increasing trend after $t = 250$\,\si{\milli\second} even for $\eta=0.2$ case. The point is, that even for $\eta=0.2$, there are still non-localized, high energy particles which can escape from the channel and reach the wells, and reflect back into the channel again. This process can explain the slightly increasing trend in the last part of the $\xi_{\text{left}}(t)$.

\begin{figure}
    \includegraphics[width=80mm]{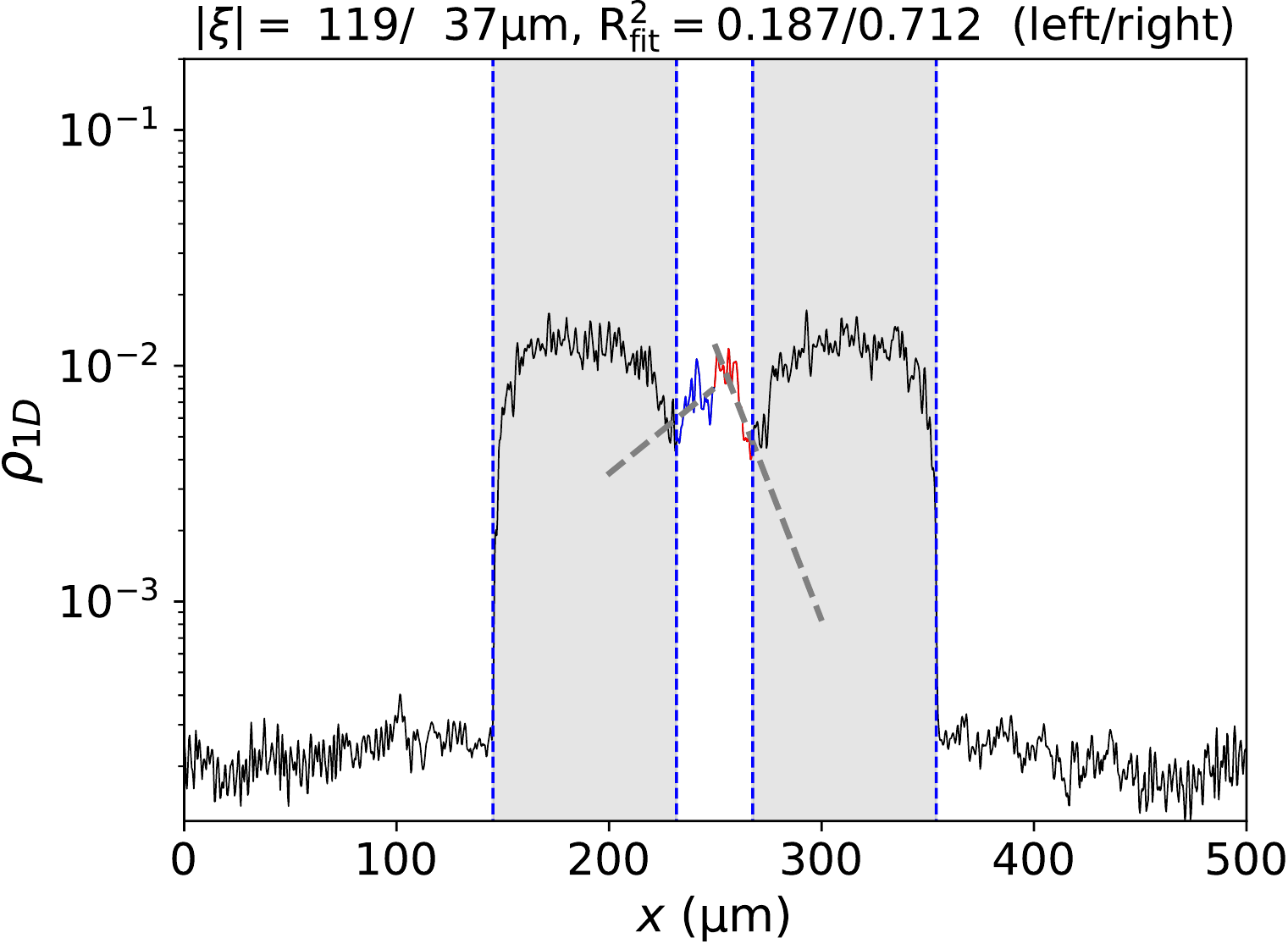}
    \caption{\label{fig:Fitting_L50}
             The one-dimensional density in a short channel is plotted at the end of the time-evolution. The left and right reservoirs are indicated with shaded areas and thin blue dashed lines. The thicker dashed lines are the linear fits to $\rho_{\text{1D}}$. The dumbbell geometry is $(L, W, R) = (36, 36, 58)$\,\si{\micro \meter}.
            }
\end{figure}

\begin{figure}
    \includegraphics[width=90mm]{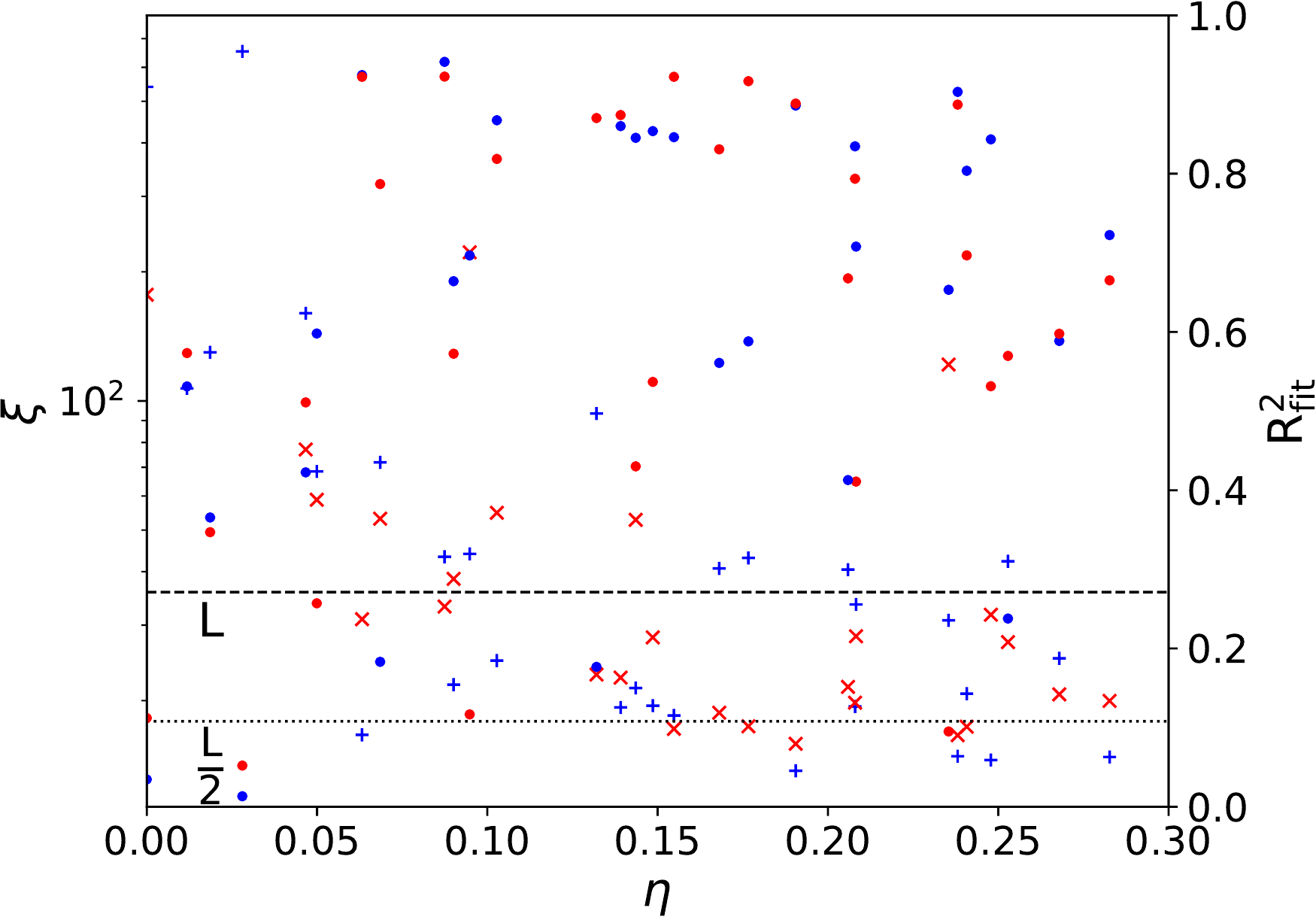}
    \caption{\label{fig:localization_vs_fillfactor_L50}
             Localization length $\xi$ as a function of $\eta$ for a dumbbell with a short channel length of 36\,\si{\micro \meter}.
            }
\end{figure}

In order to catch a localization length less than system size in a 2D system, we need to consider another important factor, the actual system size, $\sim L$. We consider a dumbbell with a short channel with $\eta=0.2$. Fig~\ref{fig:Fitting_L50} shows $\rho_{\text{1D}}$ within different segments of the dumbbells at $t > \SI{200}{\milli\second}$ for $L=$\,\SI{26}{\micro \meter}. The goodness-of-fit is low while $\xi$ is much larger than $L$. Moreover, we can not see any obvious trend in $\xi(\eta)$ or in $R_{\text{fit}}^2$ in Fig.~\ref{fig:localization_vs_fillfactor_L50}. Comparing Figs.~\ref{fig:LocalizationLength_fillfactor} and \ref{fig:localization_vs_fillfactor_L50}, we can safely conclude that in a short channel there are much less scattering events, leading to larger localization length than the actual system size. 

\begin{figure}
    \includegraphics[width=82mm]{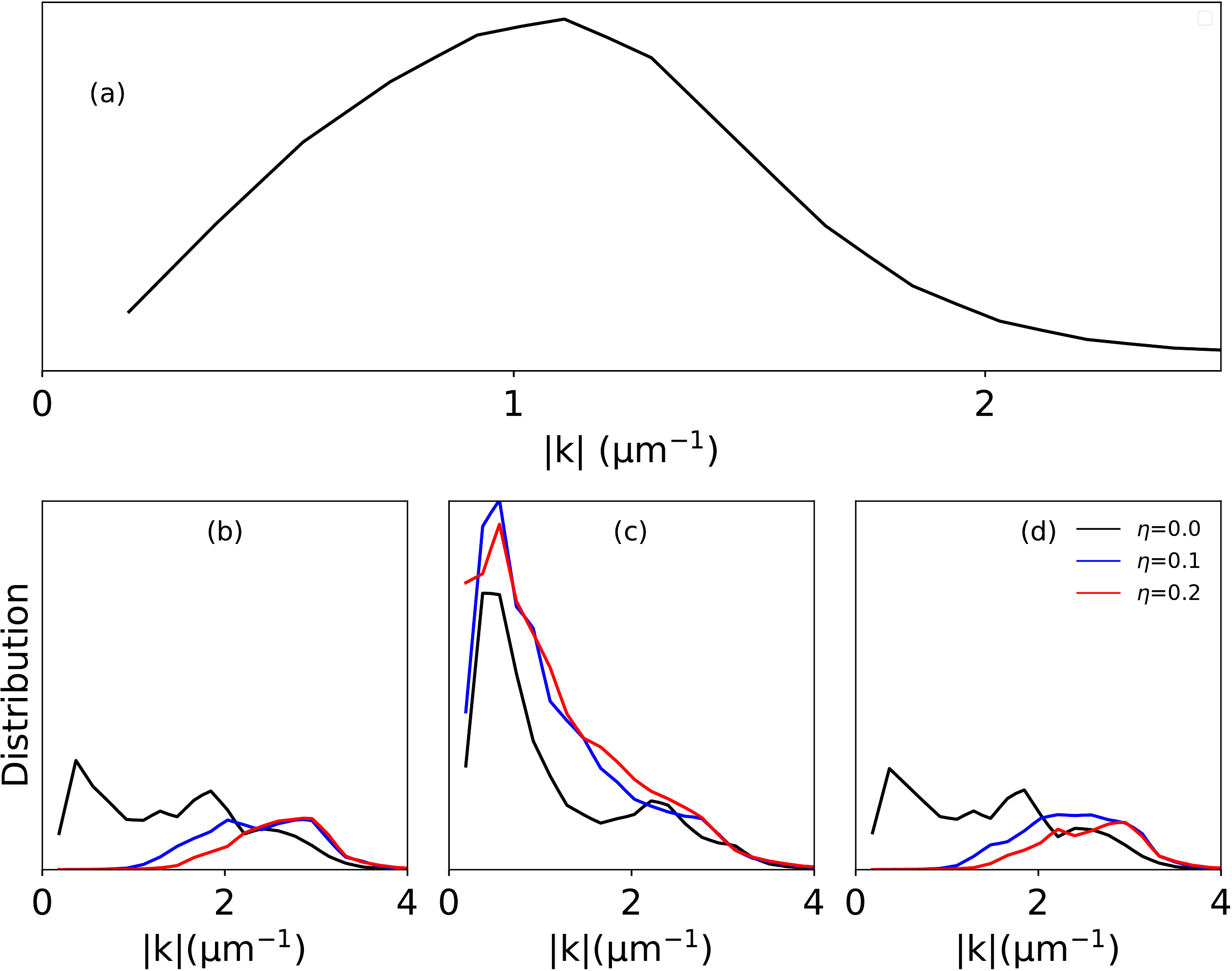}
    \caption{\label{fig:momentum_distribution}
             Momentum distribution of atoms.
             (a) depicts the initial momentum distribution of the modulus of momentum after releasing the initial harmonic trap ($t = 1$\,ms). The distribution of the modulus of momentum in the right well 
             (b) in channel and
             (c) in the left well
             (d) after $t>250$\,\si{\milli\second} is plotted.
             The dumbbell parameters are $(L, W, R) =(180, 36, 45)$\,\si{\micro\meter}.
            }
\end{figure}

Let us to look at the momentum distribution of the atoms. Figure~\ref{fig:momentum_distribution} shows $\abs{\mathbf{k}}$ in different segments of the dumbbell, derived through Fourier transform of $\rho_{\text{1D}}$. The momentum distributions of the initial wave packet at $t=\SI{1}{\milli\second}$ is shown in Figs.~\ref{fig:momentum_distribution}(a), while the momentum distributions in different regions after expansion ($t > \SI{250}{\milli\second}$) are shown in Figs.~\ref{fig:momentum_distribution}(b) to (d). Panels \ref{fig:momentum_distribution}(b) and (d) clearly show that for non-zero fill-factor the mean momentum is slightly above \SI{2}{\per\micro\meter}, indicating that particles with higher momenta escaped from the channel and reached the reservoirs. Atoms with lower momenta are trapped inside of the disordered channel as suggested by Fig.~\ref{fig:momentum_distribution}(c). In addition, Figs.~\ref{fig:momentum_distribution}(b) and (d), also points towards an energy-dependent localization, since even for the highest fill-factor ($\eta=0.2$) there are particles capable of leaving the channel. This can naively interpreted as follows: atoms with higher momenta, thus with higher kinetic energy, have shorter wavelengths and are of the order of the mean free path. The mean free-path can be approximated by the mean spacing between scatterers $\ell_{s} = \sigma/\sqrt{\eta}$, where $\sigma$ is the side-length of a single scatterer. The corresponding mean minimal distance for $\eta=0.1$ and 0.2 are $\ell_{s}=\SI{4.42}{\micro \meter}$ and \SI{3.1}{\micro \meter}, respectively. According to Fig~\ref{fig:momentum_distribution}(a) the majority of atoms have $\abs{k}=$\SI{1.11}{\per\micro \meter}, which translates to a wavelength $\lambda = 2\pi/\abs{k} = $\SI{5.6}{\micro\meter}, hence $\sigma < \ell_{s} < \lambda$. In Figs.~\ref{fig:momentum_distribution}(b) and (d), however, the distributions peak around 2--3 \si{\per\micro\meter}, with corresponding wavelengths being \SI{3.14}{\micro\meter} to \SI{2}{\micro \meter}. Therefore, $\sigma < \lambda < \ell_{s}$.

\subsection{Regularly distributed scatterers}

\begin{figure}
    \includegraphics[width=82mm]{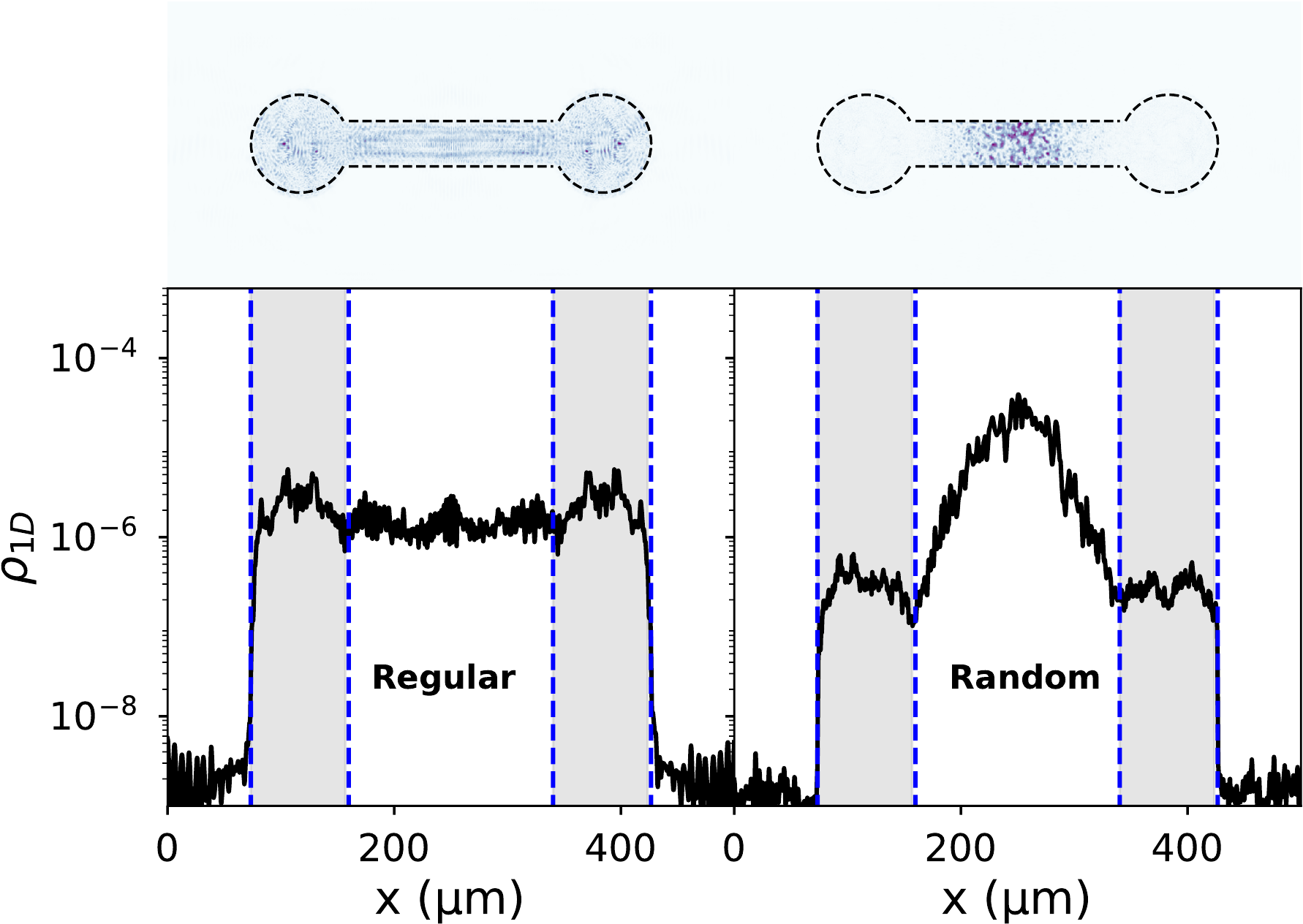}
    \caption{\label{fig:Random_Regular_Density_F0_20}
             A comparison of two-dimensional (top panels) and one-dimensional densities (bottom panels) after $t = 250$\,\si{\milli\second} time of expansion for impurities distributed regularly and randomly. The top panels provide a visual representation of $\rho_{\text{2D}}$ within the dumbbell and suggesting qualitatively different behaviour for the two cases. In case of regularly distributed impurities the density seems to be more or less uniform although it also shows weak filamentary structures, while the randomly distributed impurities seem to result in a more localized density distribution. The triangular shape of $\rho_{\text{1D}}$ on logarithmic scale is apparent for randomly distributed impurities, suggesting localisation. The left and right reservoirs are indicated with shaded areas and blue dashed lines.
            }
\end{figure}

We have also analysed the transport properties of atoms within regularly distributed scatterers, and contrasted particle transmission at $t > 250$\,\si{\milli\second} with that of random scatterers, see Fig.~\ref{fig:Random_Regular_Density_F0_20}. The bottom of Fig.~\ref{fig:Random_Regular_Density_F0_20} shows $\rho_{\text{1D}}$ at $t > \SI{250}{\milli\second}$. One can clearly see the dumbbell reservoirs being filled up by atoms for regularly distributed scatterers as atoms are in extended states while particles stay localized within the randomly disordered channel. The second obvious difference is in the bottom panels of Fig.~\ref{fig:Random_Regular_Density_F0_20} where $\rho_{\text{1D}}$ exhibits an exponentially decay profile for randomly located disorder and no decay for the regularly located potential spikes.  

\begin{figure}
    \includegraphics[width=82mm]{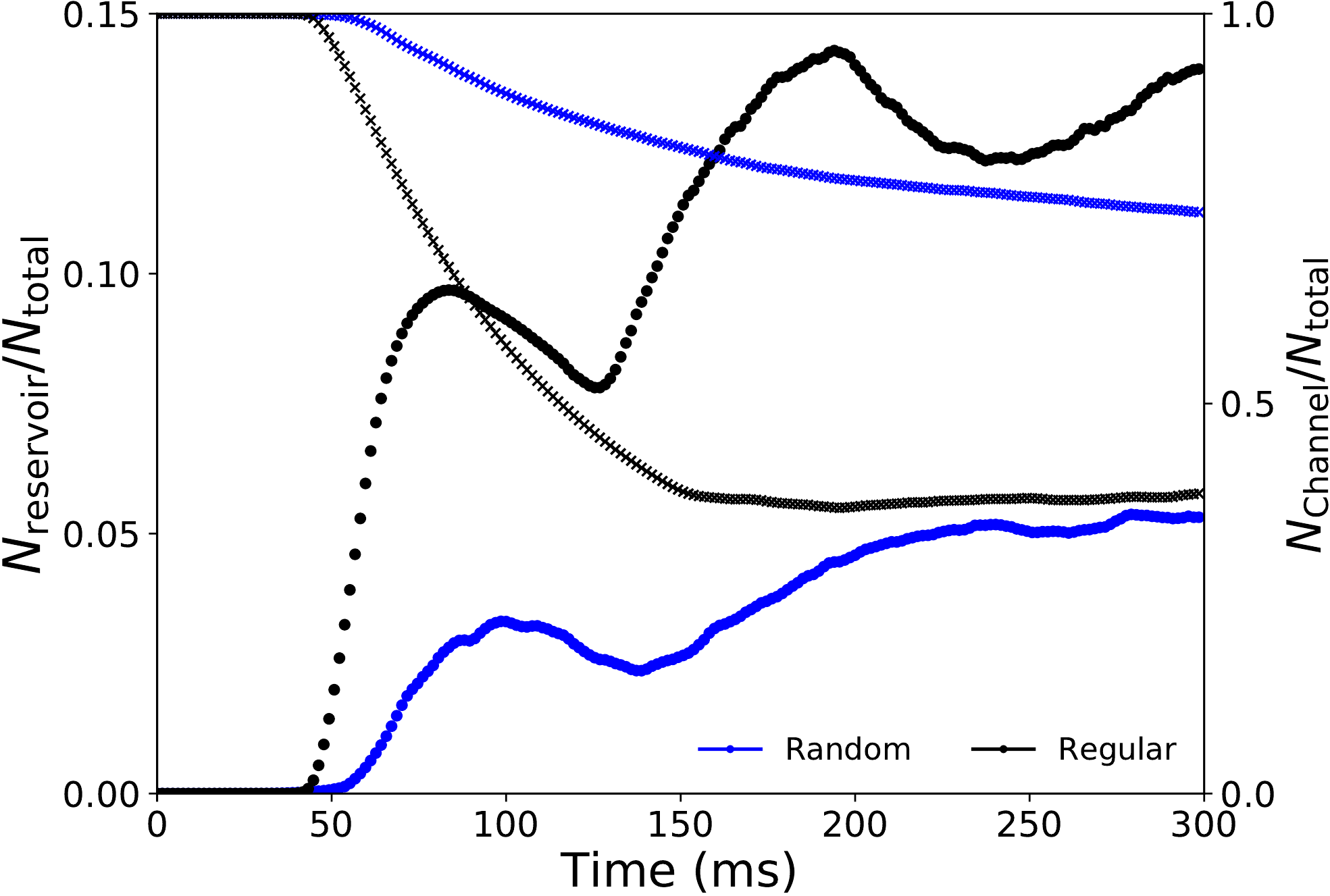}
    \caption{\label{fig:n_wl_ch_Ran_Reg}
             Normalized number of particles in the channel and in the reservoirs are depicted as functions of time.
            }    
\end{figure}

For comparison the particle numbers in the reservoirs (combined) and in the channel are shown as functions of time in Fig.~\ref{fig:n_wl_ch_Ran_Reg}. The wells of the dumbbell with regular scatterers are eventually occupied with around twice as many atoms as in the random case. In contrast, the right axes of Fig.~\ref{fig:n_wl_ch_Ran_Reg}, shows twice the number of atoms within the channel for the random system compared to that in the periodic case. The normalized atoms number in each part is derived by integrating over the density function in each segment along both horizontal and vertical axes and then normalized by the total atoms number. One may ask why the atom number in the random case does not become stable with time. The reason lies in the dynamics: some of the atoms, which are in the extended states, reach the walls of the reservoir at longer times. They then reflect back after hitting the reservoirs wall, and return into the channel, moving again toward the wells, creating a sloshing background density. This, however, does not affect our general conclusions.

\begin{figure}
    \includegraphics[width=80mm]{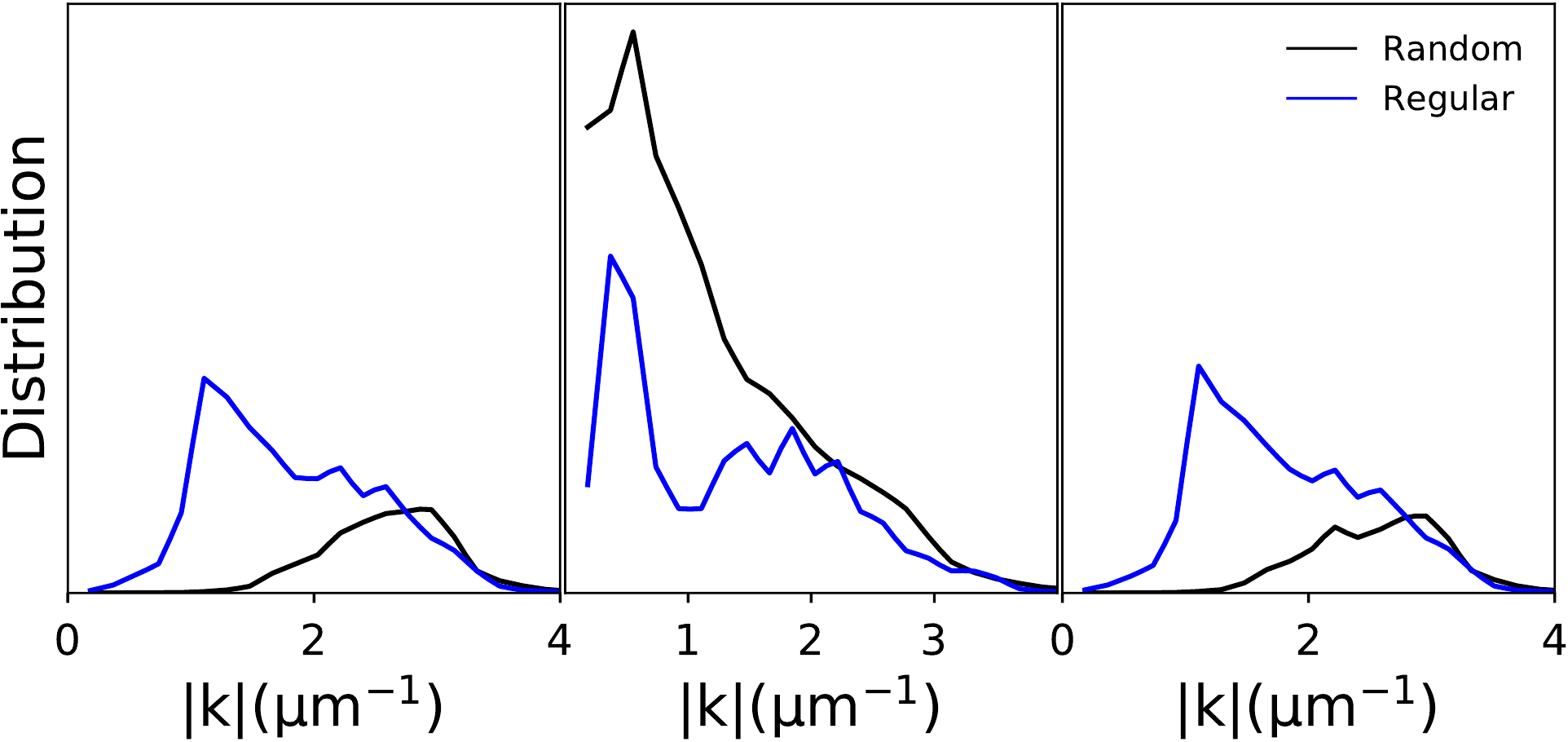}
    \caption{\label{fig:momentum_Reg_Rand_L250}
             Momentum distributions of atoms in the three segments of the dumbbell trap are shown for regularly and randomly located impurities with $\eta=0.2$. The dumbbell geometry is given by ($L, W, R) = (180, 36, 45)$\,\si{\micro\meter}.}
\end{figure}

Furthermore, we compare the momentum distributions for random and regular distributions in Fig.~\ref{fig:momentum_Reg_Rand_L250}. Atoms are in the extended states in regular case, therefore the particles distribution have a similar pattern in each segment. In contrast, particles just with high kinetic energy can reach to the left/right wells in random case, since propagation of atoms halt due to Anderson localization. Therefore, there is a significant separation between the momentum distribution in three regions of random case. The majority of particles within the channel have small $\abs{k} < 1$ while those with large $\abs{k}$, could escape from the channel and moves toward the wells. 

\subsection{Quantum impedance}

As a final measure of the  transport properties, we introduce a quantum analogue of wave impedance, $Z$, which is usually interpreted as resistance experienced by a wave propagating in a medium (cf. electronics \cite{Brillouin2003}, acoustics \cite{Dunn1986}, optics \cite{Kronig1950}). Quantum impedance was first defined by \citet{Brillouin2003} and later redefined by \citet{Khondker1988} for a typical one-dimensional scatterer, i.e., a plane wave approaching a potential barrier of finite size with the wave reflected from and transmitted through this potential barrier. For the wavefunction $\psi$, the probability current density is $j = -i \frac{\hbar}{2m} \lbrack \psi^{\ast} (\nabla \psi) - \psi (\nabla \psi^{\ast}) \rbrack$. Introducing $\phi(x) = -i \frac{\hbar}{m} \nabla \psi$, we can also write $j = \frac{1}{2} \text{Re}(\phi \psi^{\ast})$, which resembles the expression for the average power, $\frac{1}{2} \text{Re}(V I^{\ast})$, delivered in an electrical circuit. The similarity suggests the introduction of a position-dependent quantum impedance
\begin{equation*}
    Z(x) = \frac{\phi(x)}{\psi(x)}.
\end{equation*}

Specific cases, especially those considering periodic potential barriers within semiconductors, can be found in the literature \cite{Kabir1991, Ohtani1991, Griffiths1992, Sanada1994,  Sanada1994b, Sanada1995, Nelin2007, GutierrezMedina2013}. Our focus here is mainly on suppressed matter wave transport. For localized states a significant part of the probability density is confined within a small volume (relative to the available volume) and hence must have exponential decay around the boundary of this small volume. We may thus assume that $\psi(x) \sim \psi_{0}(x) e^{- \abs{x}/\xi}$, where $\psi_{0}$ is a slowly varying function and $\xi$ is a characteristic length-scale. With this assumption $Z \sim \frac{1}{\xi}$.

\begin{figure}[b]
    \centering
    \includegraphics[width=80mm]{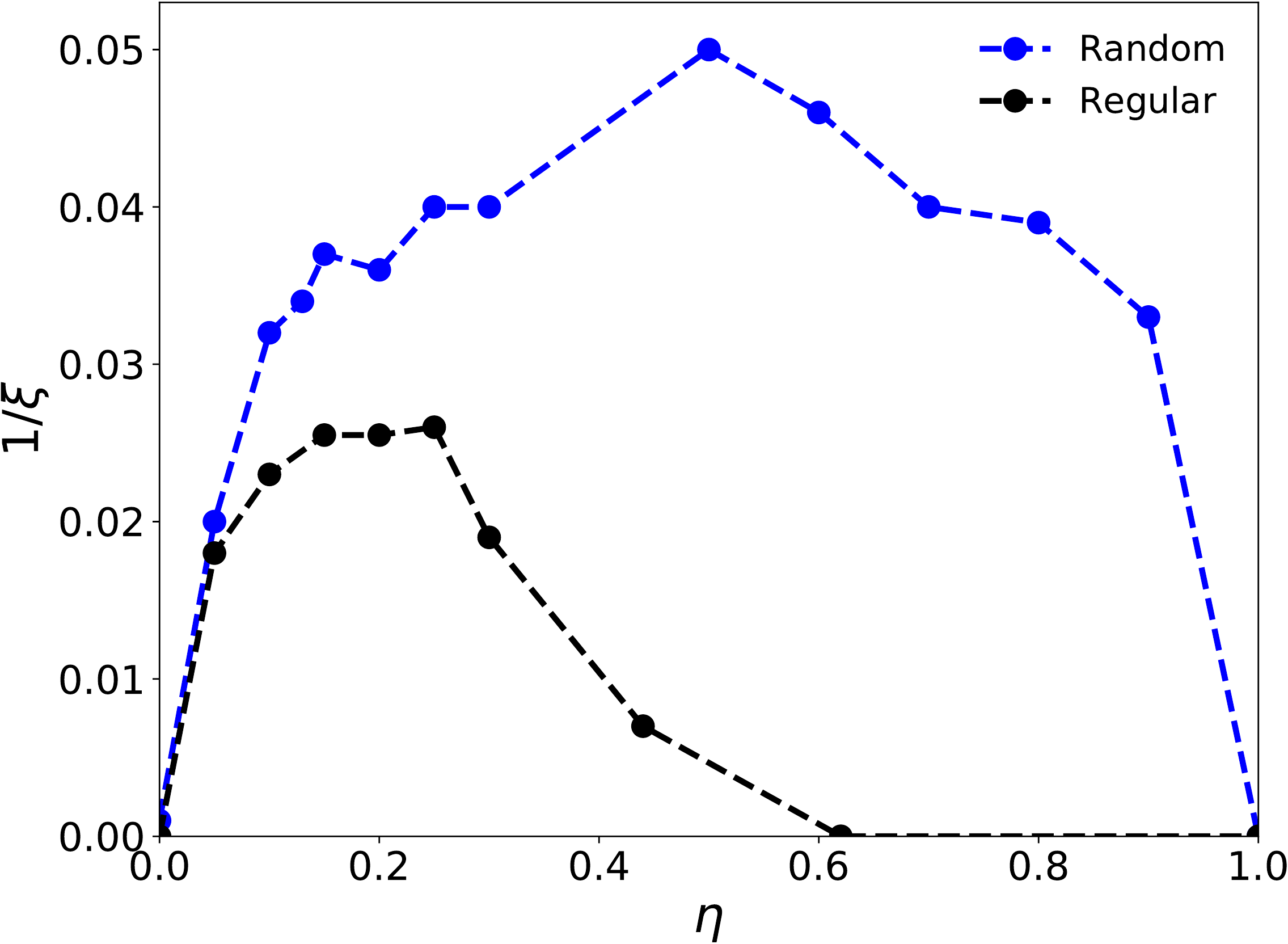}
    \caption{\label{fig:Impedance}
             The absolute value of impedance, $\abs{Z}$, is plotted as a function of $\eta$ for randomly and regularly distributed scatterers. The geometry is given by ($L, W, R) = (180, 36, 45)$\,\si{\micro\meter}.
            }    
\end{figure}

We estimate $\abs{Z}$ for a range of $\eta$ for randomly and regularly distributed impurities. The results are shown in Fig.~\ref{fig:Impedance}. As one can see for the random channel, $\abs{Z}$ increases as a function of $\eta$ and reaches its maximum for $\eta=0.5$ before it decreases again and reaches its initial value. For regular distribution, however, $\abs{Z}$ increases for $\eta \ll 1$ and reaches a lower plateau than for random disorder for $0.1 < \eta <0.2$, then rapidly diminishes for even higher fill-factors. This fast decay indicates the breakdown of our assumption for exponential decay of $\psi$, i.e., localisation itself.

\section{Conclusion}

We have studied propagation of a Bose-Einstein condensate in a 2D dumbbell-shaped trap with two types of impurities within the dumbbell channel. The dumbbell trap consists of two wells connected via a channel. The condensate is located initially in the middle of channel and propagates through channel towards the wells. We considered impurities within the channel of dumbbell distributed first randomly and then regularly. The differences between the atomic transport through these channels was investigated and showed atoms stay in localized states in a randomly located disorder channel while they are in extended states when impurities are placed periodically. We utilised the exponential decay profile of the 1D density of atoms to distinguish the localized regime from the non-localized regime. We also considered the momentum distributions of the atoms and showed particles with high energies can escape from being localized within the random disorder channel and reach the wells. These high energy particles are in extended states and demonstrate an effective mobility edge due to the finite size of the system and the particular properties of our impurity potentials. We also defined and measured an atomtronic impedance function $Z(\eta)$ for these two cases and showed a randomly located disorder channel has higher impedance in comparison with its regular counterpart. 


%

\end{document}